\documentclass[aps, pre, reprint, groupedaddress, showpacs, twocolumn]{revtex4-1}

\usepackage{amsmath,amssymb,bm}
\usepackage{color}
\usepackage{bm}
\usepackage{amsmath}
\usepackage{graphicx}
\usepackage{algorithm}
\usepackage{algpseudocode}
\usepackage{afterpage}
\usepackage{ulem}

\usepackage[dvipdfmx,
colorlinks=true,
urlcolor=blue,
citecolor=blue,
linkcolor=blue,
hyperfootnotes=false]{hyperref}

\begin{document}

\title{Corner transfer matrix renormalization group analysis of the two-dimensional dodecahedron model}

\author{Hiroshi Ueda$^{1,2}$}
\author{Kouichi Okunishi$^3$}
\author{Seiji Yunoki$^{1,4,5}$}
\author{Tomotoshi Nishino$^{6}$}

\affiliation{$^1$Computational Materials Science Research Team, 
RIKEN Center for Computational Science (R-CCS), Kobe 650-0047, Japan}
\affiliation{$^2$JST, PRESTO, Kawaguchi, 332-0012, Japan}
\affiliation{$^3$Department of Physics, Niigata University, Niigata 950-2181, Japan}
\affiliation{$^4$Computational Condensed Matter Physics Laboratory, RIKEN Cluster for 
Pioneering Research (CPR), Wako, 351-0198, Japan}
\affiliation{$^5$Computational Quantum Matter Research Team, 
RIKEN Center for Emergent Matter Science (CEMS), Wako, 351-0198, Japan}
\affiliation{$^6$Department of Physics, Graduate School of Science, Kobe University, Kobe 657-8501, Japan}

\date{\today}

\begin{abstract}

We investigate the phase transition of the dodecahedron model on the square lattice. The model 
is a discrete analogue of the classical Heisenberg model, which has continuous $O(3)$ symmetry.
In order to treat the large on-site degree of freedom $q = 20$, we develop a massively parallelized 
numerical algorithm for the corner transfer matrix renormalization group method, 
incorporating EigenExa, the high-performance parallelized eigensolver. The scaling analysis with 
respect to the cutoff dimension reveals that there is a second-order phase transition at 
$T^{~}_{\rm c}=0.4398(8)$ with the critical exponents $\nu=2.88(8)$ and $\beta=0.21(1)$. 
The central charge of the system is estimated as $c=1.99(6)$.
\end{abstract}


\maketitle

\section{Introduction}

\begin{figure}[bt]
\begin{center}
\includegraphics[width=8cm]{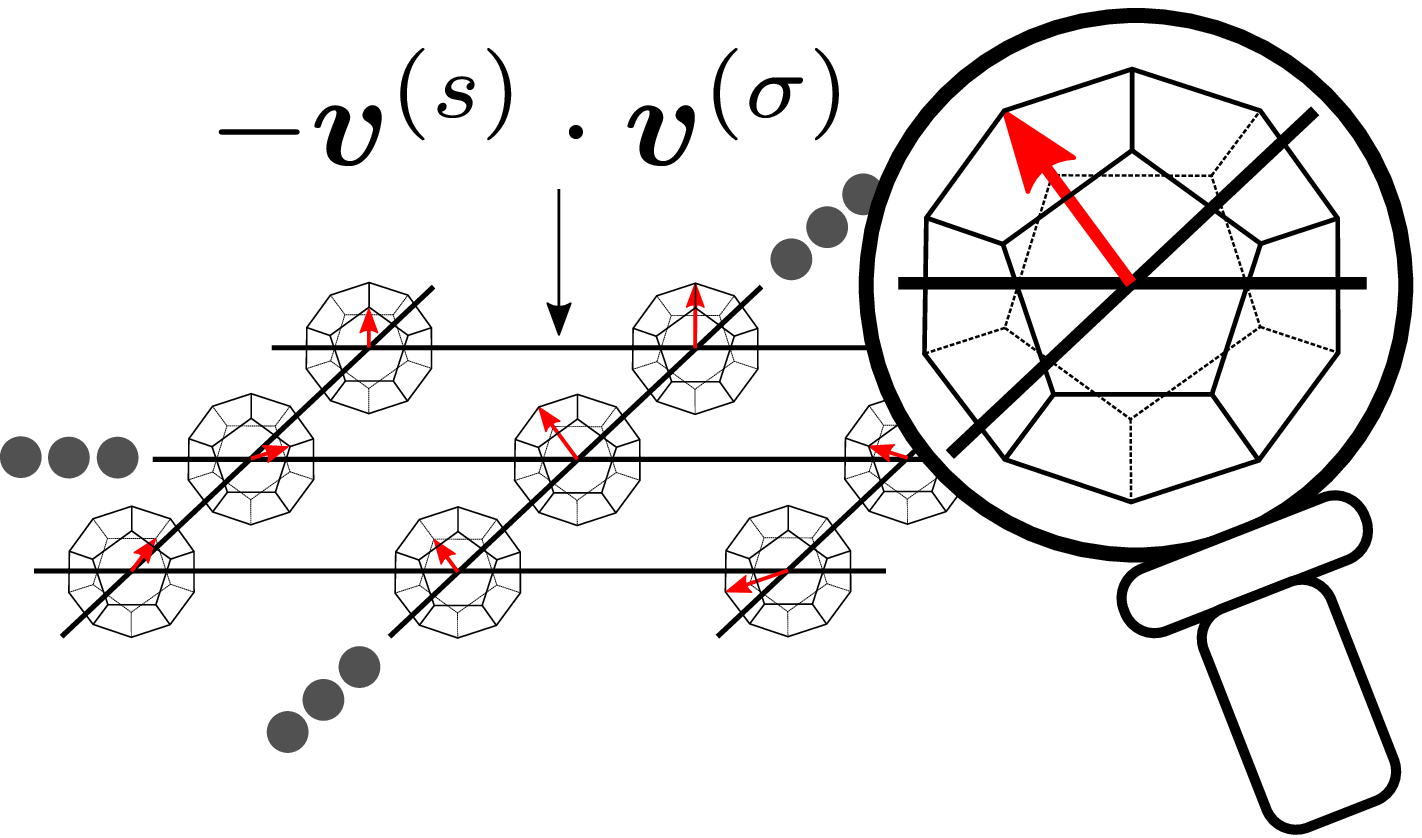}
\caption{(Color online) Dodecahedron model on the square lattice. Each unit vector spin 
points one of the 20 vertices of the dodecahedron}
\label{fig:model}
\end{center}
\end{figure}

Clarification of the role of local symmetry in phase transition is important for the fundamental 
understanding of critical phenomena. Two-dimensional (2D) polyhedron models have been 
attracting theoretical interests, in particular in their variety of phase transitions. The models
are discrete analogues of the classical Heisenberg model, which has continuous $O(3)$ symmetry. 
The polyhedron models are described by the pairwise ferromagnetic interaction  
$h^{~}_{s\sigma} = - {\bm v}^{(s)}_{~} \! \cdot {\bm v}^{(\sigma)}_{~}$ 
between neighboring sites, where 
${\bm v}^{(s)}_{~}$ with $1 \le s \leq q$ 
represents the unit-vector spin directing one of the $q$ vertices of the polyhedron. Figure~\ref{fig:model} 
shows the pictorial representation of the dodecahedron model, where $q = 20$. 

The regular polyhedron models on the square lattice have been intensively studied, and it has been 
revealed that each of them has a characteristic phase transition. 
The tetrahedron model ($q=4$) can be mapped to four-state Potts model~\cite{Wu}, and it exhibits 
second-order transition with logarithmic correction~\cite{Nauenberg, Cardy}. 
The octahedron model ($q=6$) exhibits a weak first-order phase transition~\cite{Patrascioiu, Krcmar}, 
whose latent heat is close to that of the five-state Potts model~\cite{nishino_okunishi}.
The cube model ($q=8$) can be trivially mapped to three-set of Ising models, in the same manner 
as the square model corresponds to two sets~\cite{Betts}.
Recent numerical studies on the icosahedron model ($q=12$) clarified that the model exhibits 
a continuous phase transition~\cite{Patrascioiu, Surungan,HU2017}, whose universality class may not 
be explained by the minimal unitary models in the conformal field theories (CFTs). 
Curiously, for the dodecahedron model ($q=20$), the possibility of an intermediate phase was 
suggested by Monte Carlo simulations in Refs.~[\onlinecite{Patrascioiu2}] and~[\onlinecite{Patrascioiu3}], 
whereas a single second-order transition was suggested by other Monte Carlo simulations in 
Ref.~[\onlinecite{Surungan}].
In this article, we investigate the dodecahedron model to resolve the unclear situation. 
This is a small step to answer the question how can these discrete symmetry models approximate
the classical Heisenberg model, which has no order in finite temperature~\cite{Mermin_Wagner}.

An efficient numerical method for the investigation of 2D statistical models is the corner transfer matrix 
renormalization group (CTMRG) method~\cite{ctmrg1, ctmrg2, Orus}, which is a typical tensor network 
method based on the Baxter's corner-transfer matrix (CTM) formalism~\cite{Baxter1, Baxter2, Baxter3}. 
In the CTMRG, the area of CTMs and the half of row-to-row (column-to-column) transfer matrices are iteratively extended in combination with their low-rank approximation to maintain the matrix size within a certain cutoff dimension $m$.
The  numerical accuracy of the method is  well even for small $m$, while its computational cost is proportional to $\mathcal{O}(m^3_{~}q^3_{~})$~\cite{comp_time}.
Thus, the CTMRG method enables us to obtain precise numerical data with the use of a realistic computational resource,  even for the polyhedron models with large on-site degrees of freedom. 
However, we also noted that the computational cost required for the dodecahedron model ($q=20$) is about 20 times larger than that of the icosahedron model ($q=12$). We therefore develop a massively parallelized algorithm for the CTMRG method by means of the message-passing
interface (MPI)~\cite{MPI}, combined with the numerical diagonalization package 
EigenExa~\cite{Eigenexa,web_eigenexa}, which is also MPI parallelized. 

In the previous study on the icosahedron model ($q=12$)~\cite{HU2017}, the calculations was performed 
up to $m=500$. Critical exponents associated with magnetization $M$ and correlation length $\xi$ 
are estimated by means of the finite $m$-scaling analysis~\cite{HU2017,fes1,fes2,tagliacozzo,pollmann,HU2020}.
The central charge $c$ is also extracted from the finite-$m$ scaling applied to the entanglement entropy
$S_{\rm E}^{~}$. It was suggested that the model exhibits the second-order transition with a nontrivial 
central charge $c = 1.90(2)$. Thus, a focus in the study of the dodecahedron model ($q=20$) is the 
nature of the phase transition. If it is second-order, what is the value of $c$? In this article 
we perform the finite $m$-scaling analysis for the dodecahedron model up to $m=800$.

This article is organized as follows. 
In the next section, we briefly explain the outline of the CTMRG method applied to polyhedron models. 
In Section~\ref{sec:parallel} we explain a parallelization technique implemented to the CTMRG 
method, when it is combined with EigenExa. We benchmark the numerical program on the K computer, 
which was operated in RIKEN R-CCS, through the test application on the icosahedron model.
In Section~\ref{sec:results}, we show temperature dependencies of the spontaneous magnetization and 
the entanglement entropy. We perform the finite-$m$ scaling analysis in association with the effective 
correlation length induced by the finite cutoff effect. 
The conclusions are summarized in Section~\ref{sec:summary}, and role of dodecahedral symmetry is 
discussed.

\section{Corner Transfer Matrix Formalism}
\label{sec:ctmrg}

We represent the regular polyhedron model on the square lattice in terms of the 2D tensor network, 
which is written as the contraction among 4-leg `vertex' tensors. Let us consider $q$-state vector spins 
${\bm v}^{(s)}_{~}$, ${\bm v}^{(\sigma)}_{~}$, ${\bm v}^{(s')}_{~}$, and ${\bm v}^{(\sigma')}_{~}$ of unit 
length, which are located at each corner of a unit square on the lattice. The local energy associated 
with these vector spins is written as the sum of pairwise ferromagnetic interactions
\begin{equation}
E_{s \sigma s'_{~} \sigma'_{~}}^{~} = 
h^{~}_{s\sigma}+ h^{~}_{\sigma s'} + h^{~}_{s'\sigma'} + h^{~}_{\sigma's} \, ,
\end{equation}
where $h^{~}_{s\sigma}$ denotes $- {\bm v}^{(s)}_{~} \! \cdot {\bm v}^{(\sigma)}_{~}$ as we introduced in the 
previous section. We have chosen interaction parameter as unity. The corresponding Boltzmann weight 
\begin{equation} 
W^{~}_{s \sigma s'_{~} \sigma'_{~}} = 
\exp \left[  - \frac{E_{s \sigma s'_{~} \sigma'_{~}}^{~}}{k^{~}_{\rm B} T} \right] 
\label{eq:poly2}
\end{equation}
can be regarded as the local 4-leg vertex tensor~\cite{vertex}, where $k^{~}_{\rm B}$ is the Boltzmann 
constant, and $T$ is the thermodynamic temperature. Throughout this article we choose the temperature 
scale where $k^{~}_{\rm B} = 1$. It should be noted that the vertex tensors are defined on every other unit 
squares on the lattice. The product over all the vertex tensors contained in the system represents the 
Boltzmann weight for the entire system under a specific spin configuration. Taking the configuration sum
for this weight, we obtain the partition function $Z$. 

In the CTM formalism~\cite{Baxter1, Baxter2, Baxter3}, finite-size system with square geometry is
considered. The partition function $Z$ is then represented as 
\begin{equation}
Z \equiv \mathrm{Tr} \, C^4_{~},
\label{eq:Z}
\end{equation}
where $C$ denotes the CTM corresponding to each quadrant of the finite-size system. We have used 
the fact that $C$ is real symmetric, since $W^{~}_{s \sigma s'_{~} \sigma'_{~}}$ defined in Eq.~(\ref{eq:poly2}) is 
invariant under rotation and spacial inversions of indices. In this article, we assume the ferromagnetic 
boundary condition in order to choose one of the $q$ types of the ordered state, where all the vector spins 
at the system boundary point the specified direction $s = 1$. 

In the CTMRG method~\cite{ctmrg1, ctmrg2, Orus}, we recursively update $C$ and the half row-to-row 
or half column-to-column transfer matrices $P$ toward their bulk fixed point. Thus the fixed boundary 
condition can be imposed just fixing the boundary spins in the initial transfer matrices. In order to 
prevent the exponential blow-up of the matrix dimension, these matrices are successively compressed 
by means of the truncated orthogonal transformations, which are obtained from the diagonalization of $C$. 
In this renormalization group (RG) process, the number of `kept' eigenvalues $m$ plays the role of 
the cutoff dimension~\cite{DMRG1, DMRG2}. 

After a sufficient number of iterations in the CTMRG calculation, we obtain the fixed 
point matrices ${\tilde C}$ and ${\tilde P}$, which are dependent on both $T$ and $m$. It is convenient 
to create the normalized density matrix
\begin{equation}
{\tilde \rho} \equiv \frac{ {\tilde C}^4_{~} }{ \mathrm{Tr} \, {\tilde C}^4_{~} }
\end{equation}
for the evaluation of one point functions. Spontaneous magnetization in the thermodynamic limit can be 
approximately obtained as
\begin{equation}
M(T,m) = \mathrm{Tr} \, \left[ {\bm v}^{(1)}_{~}\!\cdot{\bm v}^{(s)}_{~} \, {\tilde \rho} \right] \, ,
\end{equation}
where ${\bm v}^{(s)}_{~}$ is the vector spin located at the center of the system.
The entanglement entropy 
\begin{equation}
S_{\rm E}^{~}(T,m) = - {\rm Tr} \, {\tilde \rho}  \ln \, {\tilde \rho}
\label{eq:SE}
\end{equation}
is essential for the determination of the central charge $c$. In addition to these one point functions, we 
can calculate the effective correlation length $\xi^{~}_{\rm e}(T,m)$ by diagonalizing the renormalized 
row-to-row transfer matrices reconstructed from ${\tilde P}$. These physical functions 
are dependent on $m$, and therefore we have to take the extrapolation $m \rightarrow \infty$ by any 
means, which we consider in section~\ref{sec:results}.

\section{parallel computation}
\label{sec:parallel}

By the end of this section, we explain the massively parallelized numerical algorithm, which is 
implemented to the CTMRG method. The incorporation of the parallelized diagonalization routine 
`EigenExa'~\cite{Eigenexa} is essential in this computational programming. To the readers who 
do not care about numerics, we recommend to skip this part and proceed to the next section.

\begin{figure}[bt]
\begin{center}
\includegraphics[width=8cm]{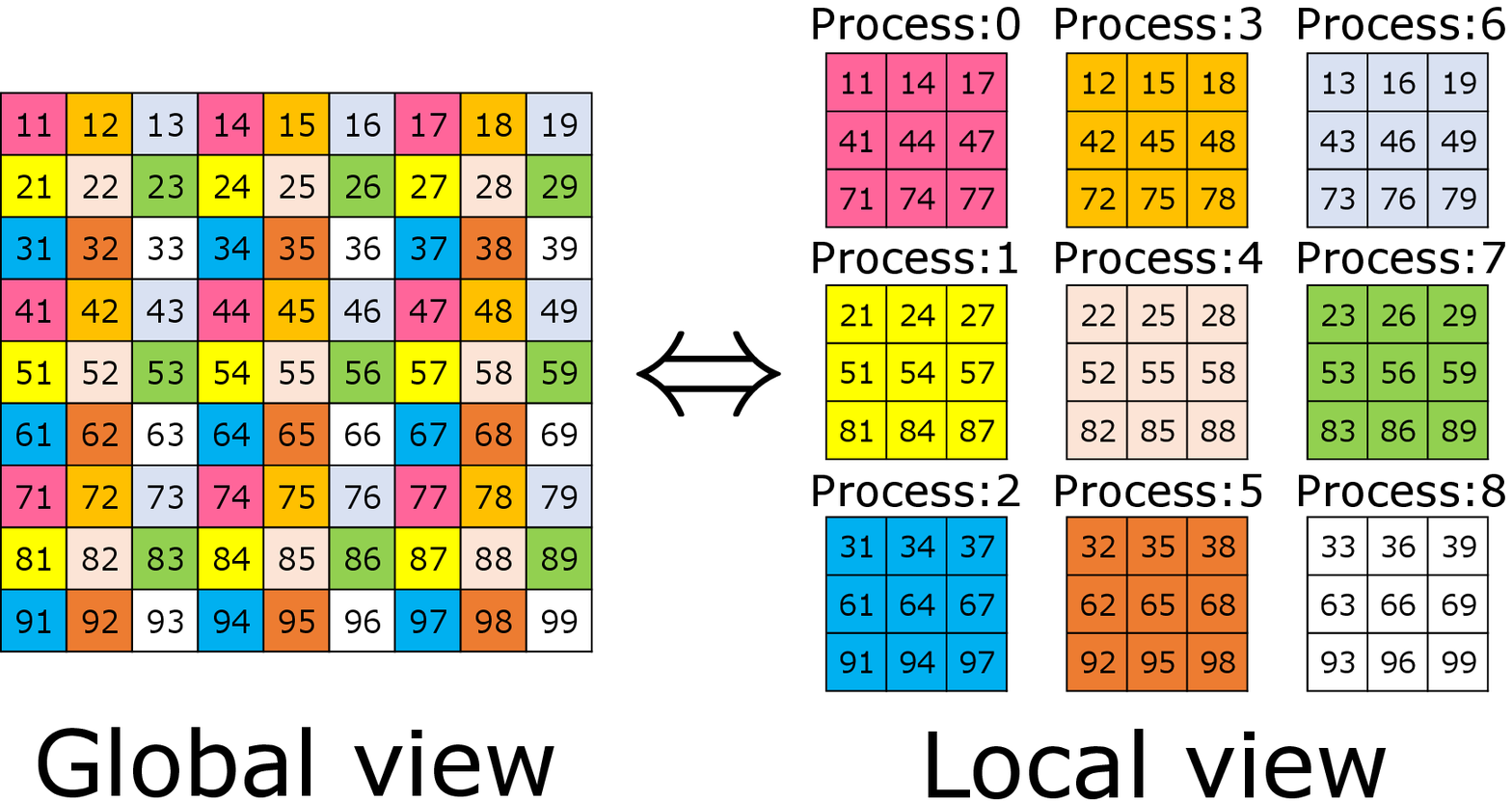}
\caption{(Color online) The $1 \times 1$ 2D block-cyclic distribution for the $9 \times 9$ matrix with 
$3\times3$ processes.}
\label{fig:one_by_one_bcd}
\end{center}
\end{figure}

Under the use of MPI~\cite{MPI}, we distribute all the elements of large-scale matrices to $n$ 
processes along ``the $1\times1$ 2D block-cyclic distribution'' shown in Fig.~\ref{fig:one_by_one_bcd}, 
where $n$ is the number of processes in MPI. We can then employ the PDGEMM routine contained 
in ``the Basic Linear Algebra Communication Subprograms'' (BLACS) package~\cite{BLACS} for 
the matrix-matrix multiplication, and can also employ EigenExa package for the diagonalization of CTMs. 
Both of these linear numerical procedures support the block-cyclic distribution. 

To achieve a high performance in matrix-matrix multiplications, we often encounter the situation where 
reordering of tensor indices is necessary. Suppose that we have a 4-leg tensor $A^{~}_{pqrs}$, 
and that we have to store the elements to another one $B^{~}_{prqs}:=A^{~}_{pqrs}$, where `$:=$' 
denotes substitution from the right to the left. This reordering can be quickly done even under the 
block-cycle distribution, as it is abbreviated in the numerical pseudocode {\bf Algorithm}~\ref{algorithm:p1}.
For the legs $p,q,r$, and $s$, respectively, we denote their leg dimension by $a, b, c$, and $d$. 
In the algorithm, the 4-leg tensor $A^{~}_{pqrs}$ is represented as a matrix $A^{~}_{i,j}$ with the 
use of combined indices $i := p+a(q-1)$ and $j := r+ c(s-1)$. Such an `addressing' is often used in the 
tensor-network frameworks. Note that the symbol {MPI\_Alltoallv} in line 2 denotes the address 
management --- to arrange which tensor elements should be stored in which array address under 
which process --- in MPI. This management enables the substitution of tensor elements in consistent 
with the block-cyclic distribution. In addition to the substitution $B^{~}_{prqs}:=A^{~}_{pqrs}$, another 
type of reordering $B^{~}_{prq}:=A^{~}_{pqr}$ between 3-leg tensors is often necessary. This process is 
represented by the pseudocode {\bf Algorithm}~\ref{algorithm:p2}. 

Generally speaking, the number of processes $n$ and the dimensions of tensor legs can vary 
during numerical calculations, therefore in principle the allocation managements should be performed 
dynamically. In the case of the CTMRG calculation, however, the maximum dimensions of all the matrices 
are always $q m$. Thus, we can make lists for the address management in advance to reduce 
communication complexity in MPI. 

\begin{figure}[bt]
\begin{algorithm}[H]
  \caption{Permutation of middle two-leg indices for a 4-leg tensor in the matrix representation.}
  \label{algorithm:p1}
   \begin{algorithmic}[1]
    \Require
    \Statex {\bf Input:} positive integer $a,b,c,d$; real ${\bf A}=\{A^{~}_{ij}\}^{1 \leq i \leq ab}_{1 \leq j \leq cd}$
    \Statex $1 \leq p \leq a$; $1 \leq q \leq b$; $1 \leq r \leq c$; $1 \leq s \leq d$
    \Ensure 
    \Statex {\bf Output:} real ${\bf B}=\{B^{~}_{\gamma \zeta}\}^{1 \leq \gamma \leq ac}_{1 \leq \zeta \leq bd}$
    \Statex
    \Statex \Comment{Matrices ${\bf A}$ and ${\bf B}$ are distributed to $n$ processes using the $1 \times 1$ 2D block-cyclic distribution.}
    \Statex
    \Function{P-Index}{${\bf A},a,b,c,d$}
    \State $\{B^{~}_{p+a(r-1),q+b(s-1)} \}:= \{A^{~}_{p+a(q-1),r+c(s-1)} \}$ 
    \Statex \Comment{Using MPI\_Alltoallv}
    \State \Return ${\bf B}$
    \EndFunction
   \end{algorithmic}
\end{algorithm}
\end{figure}
\begin{figure}[bt]
\begin{algorithm}[H]
  \caption{Permutation of the last two indices for a 3-leg tensor in the matrix representation.}
  \label{algorithm:p2}
   \begin{algorithmic}[1]
    \Require
    \Statex {\bf Input:} positive integer $a,b,c$; real ${\bf A}=\{A^{~}_{ij}\}^{1 \leq i \leq ab}_{1 \leq j \leq c}$
    \Statex $1 \leq p \leq a$; $1 \leq q \leq b$; $1 \leq r \leq c$
    \Ensure 
    \Statex {\bf Output:} real ${\bf B}=\{B^{~}_{\gamma \zeta}\}^{1 \leq \gamma \leq a}_{1 \leq \zeta \leq cb}$
    \Statex
    \Statex \Comment{Matrices ${\bf A}$ and ${\bf B}$ are distributed to $n$ processes using the $1 \times 1$ 2D block-cyclic distribution.}
    \Statex
    \Function{P-Index2}{${\bf A},a,b,c$}
    \State $\{B^{~}_{p,r+c(q-1)} \}:= \{A^{~}_{p+a(q-1),r} \}$ \Comment{Using MPI\_Alltoallv}
    \State \Return ${\bf B}$
    \EndFunction
   \end{algorithmic}
\end{algorithm}
\end{figure}

Combining the Algorithm~\ref{algorithm:p1}, \ref{algorithm:p2}, and EigenExa,  we can construct the 
CTMRG algorithm that is MPI parallelized. In {\bf Algorithm}~\ref{algorithm:ctmrg}, we present 
the resulting pseudocode for a lattice model that is invariant under $90^\circ$ rotation. The main loop 
contains four MPI\_Alltoallv communications with the cost $\mathcal{O}(m^2q^2)$, five matrix-matrix 
multiplications labeled by PDGEMM with the cost $\mathcal{O}(m^3q^2+m^2q^4)$, and the EigenExa
with the cost $\mathcal{O}(m^3q^3)$. Thus, in this algorithm, EigenExa could be the numerical bottle 
neck. Note that {\bf Algorithm}~\ref{algorithm:ctmrg} is executable on any standard computer if MPI is 
implemented, and if EigenExa is replaced by a matrix diagonalization package such as PDSYEVD 
in ScaLAPACK~\cite{ScaLAPACK}.

\begin{figure}[bt]
\begin{algorithm}[H]
  \caption{Main part of the CTMRG calculation for a vertex model with the fixed boundary condition}
  \label{algorithm:ctmrg}
   \begin{algorithmic}[1]
    \Require
    \Statex {\bf Input:} positive integer $L ,q$, and $m$; real $T$ and $\epsilon$
    \Statex $2 \leq q \leq m$ ; $0 < \epsilon \ll 1$ 
    \Statex $1 \leq i \leq q^2_{~}$; $1 \leq j \leq q^2_{~}$
    \Statex $1 \leq \alpha \leq m$ ; $1 \leq \beta \leq mq$
    \Statex $1 \leq s \leq q$; $1 \leq \sigma \leq q$
    \Ensure
    \Statex {\bf Output:} real $S^{~}_{\rm E}$ 
    \Statex $0 \leq S^{~}_{\rm E} \leq \ln m $
    \Function{Symmetric-CTMRG}{$L,q,m,T,\epsilon$}        
    \State $k:=1$; $S:=0$  \Comment{initialization}
    \State ${\bf P}=\{p^{~}_{\alpha \beta}\}$; $p^{~}_{\alpha\beta}:=\left\{ \begin{matrix} 
    1 & \alpha=\beta=1 \\ 
    0 & {\rm otherwise}
    \end{matrix} \right.$ \Comment{initialization.}
    \State ${\bm \Omega}=\{\omega^{~}_{\beta}\}$; $\omega^{~}_{\beta}:=\left\{ \begin{matrix} 
    1 & \beta=1 \\ 
    0 & {\rm otherwise}
    \end{matrix} \right.$ \Comment{initialization.}
    \State ${\bf U}=\{u^{~}_{\beta \beta}\}:=0$ \Comment{initialization.}
    \State ${\bf W}=\{w^{~}_{ij}\}$; $w^{~}_{s+q(\sigma-1),s'_{~}+q(\sigma'_{~}-1)} := W^{~}_{s \sigma s'_{~} \sigma'_{~}}$
    \Statex \Comment{local Boltzmann weight in Eq.~(\ref{eq:poly2})}
    \While{$k \leq L \land c \geq \epsilon$} \Comment{CTMRG iteration}
    \State ${\bf P}'_{~}:=\{ \omega^{~}_\alpha p^{~}_{\alpha\beta} \}$
    \State ${\bf C}:=$ {\sc Sub-CTMRG}(${\bf P},{\bf P}'_{~},{\bf W}$)
    \State $(\{u^{~}_{\gamma\gamma'}\},\{w^{~}_{\gamma}\}):=$ EigenExa($\{ c^{~}_{\gamma\gamma'}\}$) 
    \Statex \hspace{15mm} with $1 \leq \gamma \leq \min(q^k_{~},mq)$ \Comment{diagonalization}   
    \State ${\bf U}'_{~}:=$ {\sc P-Index2}($\{u^{~}_{\beta\alpha}\},m,q,m$)
    \State ${\bf X}:=$ {\sc Sub-CTMRG}(${\bf P},{\bf U}'_{~},{\bf W}$)
    \State ${\bf P}:=\{u^{~}_{\beta\alpha}\}^{\rm t}_{~} {\bf X}$ \Comment{Using PDGEMM}
    \State ${\bf P}:={\bf P}/\max_{\alpha\beta}|p^{~}_{\alpha\beta}|$ \Comment{normalization}

    \State ${\bm \Omega}:={\bm \Omega}/\sqrt[4]{\sum_\beta \omega^{4}_{\beta}}$ \Comment{normalization}
    \State $S^{~}_{\rm E}:=-\sum_{\beta} \omega^{4}_\beta \ln \omega^{4}_\beta$ \Comment{equivalent to Eq.~(\ref{eq:SE})}
    \State $c:=|1-S^{~}_{\rm E}/S|$
    \State $S:=S^{~}_{\rm E}$    
    \State $k:=k+1$
    \EndWhile
    \State \Return $S^{~}_{\rm E}$
    \EndFunction
    \Statex  
    \Function{Sub-CTMRG}{${\bf P},{\bf P}'_{~},{\bf W}$}
    \State ${\bf X}^{~}_1:={\bf P}^{\rm t}_{~}{\bf P}^{\prime}_{~}$ \Comment{Using PDGEMM}
    \State ${\bf X}^{~}_2:=$ {\sc P-Index}(${\bf X}^{~}_1,m,q,m,q$) 
    \State ${\bf X}^{~}_3:={\bf X}^{~}_2{\bf W}$ \Comment{Using PDGEMM}
    \State ${\bf X}^{~}_1:=$ {\sc P-Index}(${\bf X}^{~}_3,m,m,q,q$) 
    \State \Return ${\bf X}^{~}_1$
    \EndFunction
    \Statex \Comment{All matrices are distributed to $n$ processes using the $1 \times 1$ 2D block-cyclic distribution. Matrices ${\bf P}'_{~}$, ${\bf U}'_{~}$, ${\bf X}$, ${\bf X}^{~}_{2}$, and ${\bf X}^{~}_{3}$ are working arrays. The 2D arrays ${\bf C}$ and ${\bf X}$ can share the common physical memory in this algorithm.}
   \end{algorithmic}
\end{algorithm}
\end{figure}

\begin{figure}[tb]
\begin{center}
\includegraphics[width=8cm]{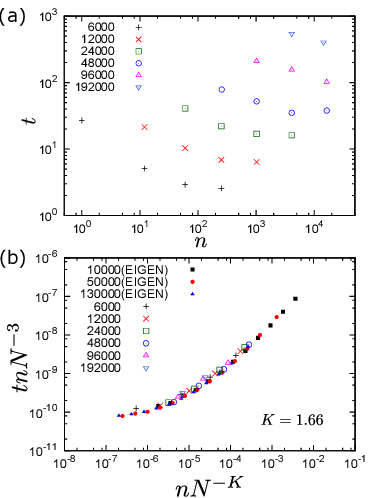}
\caption{(Color online) (a) Elapsed time (sec.) per single iteration in the parallelized CTMRG algorithm 
performed by means of K computer, when the method is applied to the icosahedron model~\cite{HU2017}. 
The holizontal axis denotes the number of nodes $n$. The maximum matrix dimension $N=mq$ is shown 
by numbers beside the legends. (b) Scaling plot for computational times required for EigenExa and 
for the time shown in Fig.~3(a).}
\label{fig:scaled_time}
\end{center}
\end{figure}

We check the performance of the {\bf Algorithm}~\ref{algorithm:ctmrg} by means of a benchmark 
computation applied to the icosahedron model ($q=12$) at the critical temperature~\cite{HU2017}.
Figure~\ref{fig:scaled_time}(a) shows the elapsed time $t$ for single iteration in the CTMRG method 
with respect to $n$, the number of nodes used, up to $n=16,380$ ($=130 \times 126$). All the 
calculations were performed on the K computer (CPU: eight-core SPARC64 VIIIfx)  installed at 
RIKEN R-CCS. If the maximum matrix dimension $N=mq$ is much larger than $n$,  the elapsed time 
decreases with respect to $n$, implying that the parallelization properly works. For $n\gtrsim N/10$, 
however,  the parallelization efficiency saturates, where the MPI communication time among the nodes 
becomes non-negligible.  

We examine a scaling hypothesis given by
\begin{equation}
t = N^{3}_{~} n^{-1}_{~} F\left( n N^{-K} \right)
\label{eq:tNn}
\end{equation}
in order to capture the relation among $t$, $N$, and $n$. The scaling function $F( y )$ has 
the asymptotic forms $F( y ) \sim y^{3/K}_{~}$ for $y \gg 1$, namely $t \sim n^{-1+3/K}$, and 
$F( y ) \sim {const.}$ for $y \to 0$. Under the ideal MPI parallelization, the exponent $K$ could be 
three, but it is empirically less than that in practical computations. For the estimation of $K$, we 
invoke the benchmark data in EigenExa with $N=1\times 10^4_{~}$, $5\times 10^4_{~}$, and 
$1.3\times 10^5_{~}$, which are available on the web page of EigenExa~\cite{web_eigenexa}.
Performing the polynomial fitting to the scaling form in Eq.~(\ref{eq:tNn}), we obtain $K=1.66$.
Assuming that the data shown in Fig.~\ref{fig:scaled_time}(a) shares the same exponent, we 
show the scaling plot for all the bench-mark data in Fig.~\ref{fig:scaled_time}(b). 
The plotted points almost collapse on a certain scaling curve, and the result supports the fact that 
the diagonalization of CTMs by EigenExa is certainly the numerical bottleneck.

\section{scaling analysis}
\label{sec:results}

\begin{figure}[tb]
\begin{center}
\includegraphics[width=8cm]{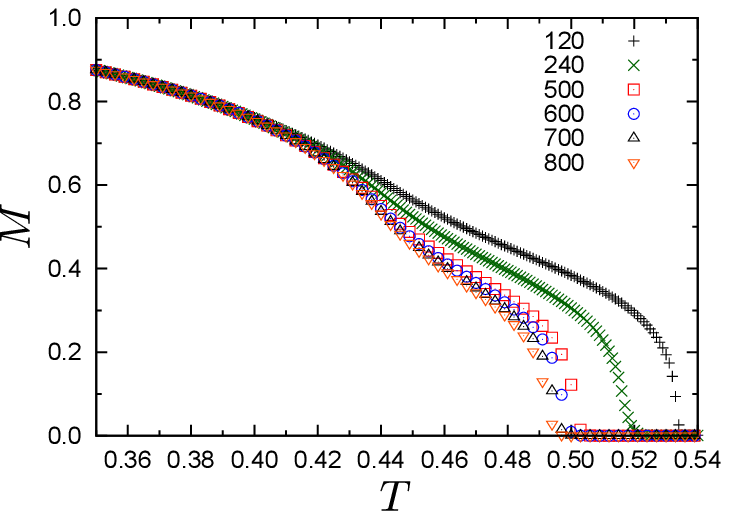}
\caption{(Color online) Temperature dependence of spontaneous magnetization $M(T,m)$.}
\label{fig:mag_vs_t}
\end{center}
\end{figure}

We performed the CTMRG calculation for the dodecahedron model, assuming the ferromagnetic 
boundary conditions. We choose the cutoff dimensions up to $m=800$~\cite{request} for all the numerical data analyses shown in this section. Figure~\ref{fig:mag_vs_t}
shows the temperature dependence of the spontaneous magnetization $M(T,m)$. The overall 
behavior of the magnetization, which exhibits a shoulder-like structure in the region 
$0.45 \lesssim T\lesssim 0.5$, is very similar to $M(T,m)$ observed in the icosahedron model~\cite{HU2017}.

We perform the finite-$m$ scaling analysis~\cite{HU2017,fes1,fes2,tagliacozzo,pollmann,HU2020}, in order to 
check whether the transition is second-order or not. At the fixed point --- the large system size limit --- 
of the CTMRG method, the presence of finite cutoff dimension $m$ modifies the intrinsic correlation 
length $\xi(T)$ to an effective one $\xi^{~}_{\rm e}(T,m)$. At the critical temperature $T = T_{\rm c}^{~}$ 
the behavior
$\xi^{~}_{\rm e}(T_{\rm c}^{~},m) \sim m^\kappa_{~}$ 
is expected, where $\kappa$ is a particular exponent~\cite{fes1,fes2,tagliacozzo}. Meanwhile, the 
intrinsic correlation length $\xi(T)$ away from the critical point obeys $\xi(T) \sim |T-T^{~}_{\rm c}|^\nu_{~}$, 
where $\nu$ is the exponent characterizing the divergence of the correlation length. Taking account 
of these relations, we can assume the finite-$m$ scaling form 
\begin{align}
\xi^{~}_{\rm e}(T,m) \sim m^{\kappa}_{~} f\left((T-T^{~}_{\rm c})m^{\kappa /\nu}_{~} \right)\,,
\label{FSS_xi}
\end{align}
where the scaling function behaves as $f( y ) \sim | y |^{- \nu}_{~}$ for $y \gg 1$ and $f( y ) \sim {const.}$ 
for $y \to 0$. We can also assume the finite-$m$ scaling form 
\begin{align}
M(T,m) \sim m^{-\kappa \beta/\nu}_{~} g\left((T-T^{~}_{\rm c})m^{\kappa /\nu}_{~} \right) 
\label{FSS}
\end{align}
for the spontaneous  magnetization, where  $\beta$ denotes the critical exponent for the magnetization, 
and $g$ is a scaling function. It should be noted that Eqs. (\ref{FSS_xi}) and (\ref{FSS}) are basically 
equivalent to the conventional finite-size scalings if we substitute the system size $\ell$ to 
$\xi_{\rm e}(T_{\rm c}^{~},m) \sim m^\kappa$. For the bipartite entanglement entropy, the 
finite-size scaling form 
$S^{~}_{\rm E}(T_{\rm c}^{~},\ell) \sim \frac{c}{6} \log \ell + const.$ 
suggests that the effective scaling dimension for $e^{S_{\rm E}(T_{\rm c}^{~},m)}_{~}$ can be 
expressed as $c/6$~\cite{Vidal,Calabrese}. Thus, we can assume the finite-$m$ scaling form 
\begin{align}
e^{S_{\rm E}^{~}(T,m)}_{~} \sim m^{c \kappa / 6}_{~} \, h\left( (T-T^{~}_{\rm c})m^{\kappa / \nu}_{~} \right) 
\label{eq:scaled_ee}
\end{align}
for the entanglement entropy, where the scaling function behaves as 
$h( y ) \sim | y |^{- \nu}_{~}$ for $y \gg 1$ and $h( y ) \sim {const.}$ for $y \to 0$. 

In order to estimate scaling parameters, we employ the Bayesian scaling analysis proposed in Ref. [\onlinecite{Harada,Harada2}], which is based on the Gaussian process regression for a smooth scaling function.
 We perform the Bayesian fitting of the scaling parameters with varying a range of $T$ and $m$ in input data to determine estimation errors.
Moreover, we check the stability of the resulting parameters against corrections to scaling in Appendix~\ref{appendix}.
In the following, we basically present the final results of the scaling parameters in Eqs.~(\ref{FSS_xi}), (\ref{FSS}), and (\ref{eq:scaled_ee}).

\begin{figure}[tb]
\begin{center}
\includegraphics[width=8cm]{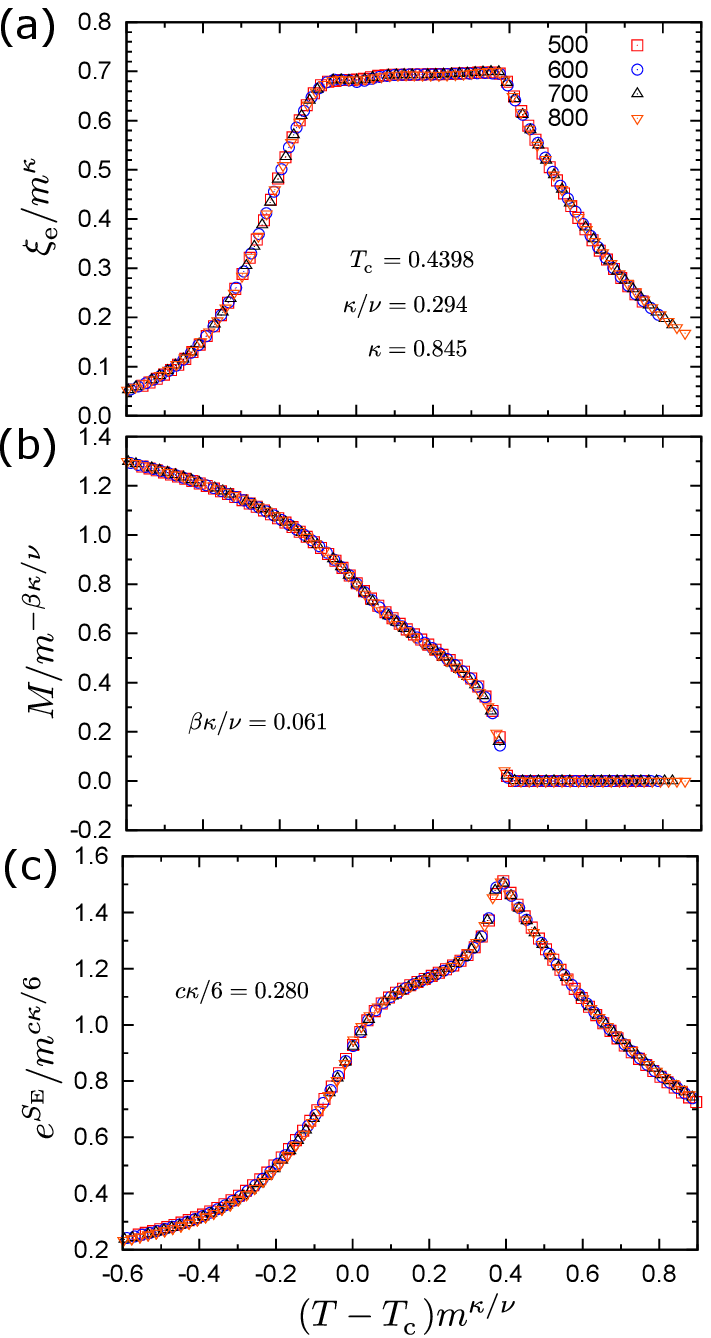}
\caption{(Color online) Scaling plots for 
(a) effective correlation length $\xi^{~}_{\rm e}(T,m)$ [Eq.~(\ref{FSS_xi})],  
(b) magnetization $M(T,m)$ [Eq.~(\ref{FSS})] , 
and (c) the exponential of the entanglement entropy $e^{S_{\rm E}(T,m)}_{~}$ [Eq.~(\ref{eq:scaled_ee})].
Note that correction terms to scaling in Appendix A are not included in these scaling plots.
}
\label{fig:scaled}
\end{center}
\end{figure}

We empirically find that the analysis on $\xi^{~}_{\rm e}(T,m)$ is more stable than that for $M(T,m)$ and $e^{S_{\rm E}^{~}(T,m)}_{~}$. 
From the calculated $\xi^{~}_{\rm e}(T,m)$ in the temperature range $0.35 \le T \le 0.56$, the values $T^{~}_{\rm c}=0.4398(8)$, $\nu=2.88(8)$ and $\kappa=0.845(4)$ are extracted. 
Figure~\ref{fig:scaled}(a) shows the corresponding scaling plot for $\xi^{~}_{\rm e}(T,m)$, where the data well collapse to a scaling function, which exhibit an intermediate plateau, as it was observed in the icosahedron model~\cite{HU2017}.

Using the obtained $T^{~}_{\rm c}$, $\nu$ and $\kappa$, we can further estimate $\beta=0.21(1)$ 
by means of the Bayesian analysis applied to $M(T,m)$ shown in Fig.~\ref{fig:mag_vs_t}. The resulting 
scaling plot is presented in Fig.~\ref{fig:scaled}(b), where the scaling function exhibits the shoulder 
structure. We finally perform the Bayesian analysis for $e^{S_{\rm E}(T,m)}_{~}$, and estimated the 
value of central charge $c=1.99(6)$. The scaling plot in Fig.~\ref{fig:scaled}(c) clearly shows that the 
calculated $e^{S_{\rm E}(T,m)}_{~}$ also collapsed on a scaling function, which exhibits a nontrivial 
intermediate structure.

It should be noted that for a 2D classical system at criticality, the central charge $c$ and $\kappa$ can 
be related with each other through the nontrivial relation,
\begin{equation}
c\kappa/6 = \left(1+\sqrt{12/c} \right)^{-1}_{~}, 
\label{eq:c_kappa}
\end{equation}
which was originally derived from the matrix-product-state description of 1D critical quantum systems~\cite{pollmann}. 
The relation is satisfied within the error bars if the above estimations of $c = 1.99$ and $\kappa=0.845$ are substituted. 
This fact provides a complemental check to the present finite-$m$ scaling analysis performed to the numerically calculated results. 
Since the estimated values of the exponents in the dodecahedron model are different from $\nu=1.62(2)$ and $\beta=0.12(1)$in the icosahedron model~\cite{HU2017}, phase transitions of these two models belong to different universality classes.

\section{Summary and discussion}
\label{sec:summary}

We have investigated the phase transition and critical properties of the dodecahedron model on the 
square lattice, where the vector spin has twenty degrees of freedom ($q=20$). In order to deal with 
the large on-site degree of freedom, we developed the massively parallelized CTMRG algorithm 
cooperating with the EigenExa~\cite{Eigenexa,web_eigenexa}. Spontaneous agnetization $M(T,m)$, 
effective correlation length $\xi_{\rm e}^{~}(T,m)$, and entanglement entropy $S_{\rm E}^{~}(T,m)$ 
are calculated for the cutoff dimensions $m$ up to $800$. The finite-$m$ scaling 
analyses~\cite{fes1,fes2,tagliacozzo,pollmann,HU2017} around the transition temperature revealed 
that the model undergoes a single second-order phase transition at $T^{~}_{\rm c}=0.4398(8)$, 
which is consistent with the Monte Carlo simulations in Ref.~[\onlinecite{Surungan}]. We also estimated 
the scaling exponents $\nu=2.88(8)$, $\beta=0.21(1)$, and the central charge $c=1.99(6)$. 

\begin{figure}[tb]
\begin{center}
\includegraphics[width=8cm]{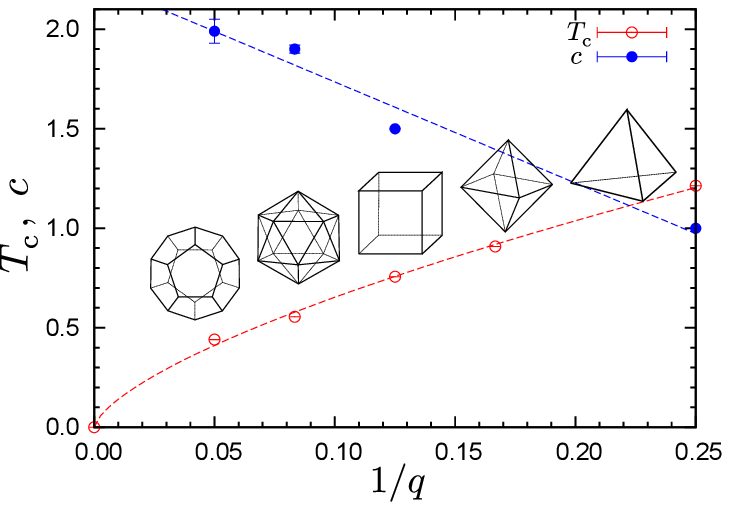}
\caption{(Color online) Critical temperatures and central charges in the regular polyhedron models. 
Broken lines are guides for eyes.}
\label{fig:data_poly_tc}
\end{center}
\end{figure}

Let us summarize the critical temperatures and central charges for the series of regular polyhedron 
models in Fig.~\ref{fig:data_poly_tc}. The transition temperature monotonically decreases with respect 
to the number of on-site degree of freedom $q$. The behavior in $T_{\rm c}^{~}$ is consistent with 
the fact that it converges to zero in the large-$q$ limit, which is the classical Heisenberg 
model~\cite{Mermin_Wagner}. Note that the octahedron model ($q=6$) is known to exhibit a weak 
first-order phase transition. Meanwhile, the central charge monotonically increases with $q$.
The exact value $c = 1$ is known for the tetrahedron model ($q=4$),  which corresponds to four-state 
Potts model. Also for the cubic model ($q=8$), which is nothing but three set of  Ising models, the value 
$c = 3/2$ is known.

So far, we have no theoretical explanation for the central charges $c = 1.90(2)$ and $c=1.99(6)$, respectively, for the icosahedron model and dodecahedron model. How can we explain the universality 
classes of the phase transitions, and can interpret the intermediate shoulder structures in the scaling 
functions? In these two models, there are several ways of introducing anisotropy to the vector spins, 
according to the subgroup structure of the polyhedral symmetry~\cite{symmetry}. A preliminary numerical 
calculation suggest that introduction of XY anisotropy to these models induce KT transitions. A more promising 
deformation is the introduction of the cubic anisotropy. If the phase transition splits into two different
ones subject to different subgroup symmetries, the value of central charge in each transition would
explain the value of $c$ obtained in this study. Complementary, an effective field theoretical treatment 
within the regular polyhedron symmetry is also a non-trivial future problem.

If we consider polyhedron models in general, in addition to the regular ones, semi-regular (or truncated) 
polyhedron model would be important candidates for the future study of attacking the large-$q$ limit. 
The pioneering work by Kr\v{c}m\'ar, et al shows that truncated tetrahedron model exhibit two phase 
transitions~\cite{Krcmar}. If we introduce the truncation scheme to the current study, we have to treat the 
truncated icosahedron, which has 60 on-site degrees of freedom. In a couple of years realistic computation 
will be possible for this system. At present, rhombic icosahedron model ($q=24$) can be the next target 
of the analysis in near future.

\acknowledgments
H.U. thanks Y.~Hirota and T.~Imamura for helpful comments on the EigenExa and S.~Morita for 
discussions of the MPI parallelization. The work was partially supported by KAKENHI 
No. 26400387, 17H02926, 17H02931, and 17K14359, and by JST PRESTO No. JPMJPR1911, and 
by MEXT as ``Challenging Research on Post-K computer'' (Challenge of Basic Science: 
Exploring the Extremes through Multi-Physics Multi-Scale Simulations). This research used 
computational resources of the K computer provided by the RIKEN R-CCS through the HPCI System 
Research project (Project ID:hp160262) and of the HOKUSAI-Great Wave supercomputing system at RIKEN.

\appendix
\section{Corrections to scalings and  their $m$ dependences}
\label{appendix}

\begin{table}[b]
  \begin{center}
    \caption{Transition temperatures and caling exponents estimated by Eqs.~(\ref{FSS_xi})--(\ref{eq:scaled_ee}) 
    and Eqs.~(\ref{FSS_xi_wc})--(\ref{eq:scaled_ee_wc}) from the data sets A: $m \in \{120,240,500,800\}$ and 
    B: $m\in\{500,600,700,800\}$.}
    \label{table:scaling_parameters}
    \begin{tabular}{ccccccc} \hline
      Set & Scaling Eqs. & $T^{~}_{\rm c}$ & $\kappa$ & $\nu$ & $\beta$ & $c$ \\ \hline \hline
      A & (8)-(10) & $0.4406(2)$ & $0.858(1)$ & $2.92(2)$ & $ 0.22(1) $ & $1.90(1)$ \\ \hline
      B & (8)-(10) & $0.4404(2)$ & $0.842(1)$ & $2.92(1)$ & $0.21(1)$ & $1.96(2)$ \\ \hline
      A & (A1)-(A3) & $0.4408(4)$ & $0.844(3)$ & $2.64(5)$ & $ 0.21(1) $ & $ 1.99(3) $ \\ \hline
      B & (A1)-(A3) & $0.4397(7)$ & $0.845(4)$ & $2.86(6)$ & $0.21(1)$ & $2.00(4)$ \\ \hline
\\ \hline      
Set & Scaling Eqs. & $\omega_1$ & $\omega_2$ & $\omega_3$ \\ \hline \hline
A & (A1)-(A3) & $0.8(1)$ & $ 1.7(1) $ & $ 0.4(1) $ \\ \hline
B & (A1)-(A3) & $0.34(2)$ & $1.4(2)$ & $0.7(1)$ \\ \hline
    \end{tabular}
  \end{center}
\end{table}
We present details of the finite-$m$ scaling for the CTMRG results of the dodecahedron model. 
As mentioned in the main text, a CFT describing the universality class of the dodecahedron 
model is not specified yet. 
Thus, it is difficult to directly estimate how the fitting for the leading scaling functions of Eqs.~(\ref{FSS_xi}), (\ref{FSS}), and (\ref{eq:scaled_ee}) is stable against correction terms associated with less relevant scaling dimensions.
Thus, replacing the system size $L$ with $m^{\kappa}$ in the standard finite-size scaling with corrections, we phenomenologically introduce the finite-$m$ scaling functions with correction terms as follows, 
\begin{align}
\xi_{\rm e}^{~}( T, m ) & \sim  ~ m_{~}^{\kappa} \, 
\Big[ \, f\left( ( T - T_{\rm c}^{~} ) m^{\kappa /\nu}_{~} \right) \nonumber \\ 
& ~~ + m^{ - \kappa\omega^{~}_1 } f_1^{~} \! \left( ( T - T_{\rm c}^{~} )m_{~}^{\kappa /\nu} \right) \, \Big] \, ,
\label{FSS_xi_wc} \\
M(T,m) & \sim  m^{-\kappa\beta/\nu}_{~} \Big[  g\left((T-T^{~}_{\rm c})m^{\kappa /\nu}_{~} \right) \nonumber \\ 
& ~~ + m^{-\kappa\omega^{~}_2} \, g^{~}_1 \! \left((T-T^{~}_{\rm c})m^{\kappa /\nu}_{~} \right) \, \Big] \, ,
\label{FSS_wc} \\
e^{S_{\rm E}^{~}(T,m)}_{~} & \sim ~~ m^{c\kappa/6}_{~} \Big[ \, h\left((T-T^{~}_{\rm c})m^{\kappa /\nu}_{~} \right) \nonumber \\ 
& ~~ + m^{-\kappa\omega^{~}_3} \, h^{~}_1 \! \left((T-T^{~}_{\rm c})m^{\kappa /\nu}_{~} \right) \, \Big] 
\label{eq:scaled_ee_wc}
\end{align}
where $f^{~}_1$, $g^{~}_1$, and $h^{~}_1$ denote scaling functions for correction terms and $\omega^{~}_1$,  $\omega^{~}_2$,  and $\omega^{~}_3$ are irrelevant exponents.

Let us evaluate the leading scaling parameters in the dodecahedron model by comparing Bayesian scaling analyses~\cite{Harada,Harada2} for Eqs.~(\ref{FSS_xi})--(\ref{eq:scaled_ee}) and those for Eqs.~(\ref{FSS_xi_wc})--(\ref{eq:scaled_ee_wc}) including the correction terms.
Here, It should be noted that the fitting results may depend on the range of cut-off dimension $m$. 
To check the $m$-dependence, we use two sets of data: one is the set A, $m \in \{120,240,500,800\}$, 
which contains small $m$ cases, and the other is the set B, $m\in\{500,600,700,800\}$. 

Table~\ref{table:scaling_parameters} summarize the result of numerical fitting analysis. Since the data set A 
contains small $m$ cases, the estimated $\kappa$ and $c$ from Eqs.~(\ref{FSS_xi})--(\ref{eq:scaled_ee}) 
and $\nu$ from Eqs.~(\ref{FSS_xi_wc})--(\ref{eq:scaled_ee_wc}) show relatively large deviation.
Meanwhile, the transition temperature $T^{~}_{\rm c}$ and the exponents $\kappa, \nu, \beta$ and $c$ obtained from the data set B are consistent both for the the scaling functions with and without correction terms. 
Thus, $m$ in the data set B are sufficiently large for the estimation of these values, although the irrelevant exponents $\omega^{~}_1$, $\omega^{~}_2$, and $\omega^{~}_3$ exhibit large $m$ dependencies. 
Discarding the scaling result from the data set A, we obtain the values $T^{~}_{\rm c}=0.4398(8)$, 
$\kappa=0.845(4)$, $\nu=2.88(8)$, $\beta=0.21(1)$, $c=1.99(6)$, which were presented in the main text. 
We have determined error bars of the final estimation of exponents so as to include the error bars of the fitting results for the data set B.
Indeed, the scaling plot using the determined exponents in Fig.~\ref{fig:scaled} well collapses to scaling curves.


\begin{thebibliography}{99}
\bibitem{comp_time} In general,  a larger cutoff dimension $m$ is also required for a CTMRG calculation 
with a large $q$ model.
\bibitem{request} We requested computational time around a day with $24 \times 24$ nodes 
on the K computer, where each node has a process.
\bibitem{vertex} The term `vertex' used here corresponds to a square on the diagonal square lattice, not 
to a vertex of the polyhedron.
\end{thebibliography}
\end{document}